\documentstyle[epsf,graphics,general_cite]{mn}
\newif\ifAMStwofonts
\def\h{$^{\rm h}$}
\def\m{$^{\rm m}$}

\ifoldfss
  \ifCUPmtlplainloaded \else
    \NewTextAlphabet{textbfit} {cmbxti10} {}
    \NewTextAlphabet{textbfss} {cmssbx10} {}
    \NewMathAlphabet{mathbfit} {cmbxti10} {} % for math mode
    \NewMathAlphabet{mathbfss} {cmssbx10} {} %  "   "    "
  \fi
  \ifAMStwofonts
    \ifCUPmtlplainloaded \else
      \NewSymbolFont{upmath} {eurm10}
      \NewSymbolFont{AMSa} {msam10}
      \NewMathSymbol{\upi}     {0}{upmath}{19}
      \NewMathSymbol{\umu}     {0}{upmath}{16}
      \NewMathSymbol{\upartial}{0}{upmath}{40}
      \NewMathSymbol{\leqslant}{3}{AMSa}{36}
      \NewMathSymbol{\geqslant}{3}{AMSa}{3E}

       \let\le=\leqslant
       
    \fi
  \fi
\fi % End of OFSS

\ifnfssone
  \newmathalphabet{\mathit}
  \addtoversion{normal}{\mathit}{cmr}{m}{it}
  \addtoversion{bold}{\mathit}{cmr}{bx}{it}
  \newmathalphabet{\mathbfit} % math mode version of \textbfit{..}
  \addtoversion{normal}{\mathbfit}{cmr}{bx}{it}
  \addtoversion{bold}{\mathbfit}{cmr}{bx}{it}
  \newmathalphabet{\mathbfss} % math mode version of \textbfss{..}
  \addtoversion{normal}{\mathbfss}{cmss}{bx}{n}
  \addtoversion{bold}{\mathbfss}{cmss}{bx}{n}
  \ifAMStwofonts
    \ifCUPmtlplainloaded \else
      \UseAMStwoboldmath
      \makeatletter
      \new@mathgroup\upmath@group
      \define@mathgroup\mv@normal\upmath@group{eur}{m}{n}
      \define@mathgroup\mv@bold\upmath@group{eur}{b}{n}
      \edef\UPM{\hexnumber\upmath@group}
      \new@mathgroup\amsa@group
      \define@mathgroup\mv@normal\amsa@group{msa}{m}{n}
      \define@mathgroup\mv@bold\amsa@group{msa}{m}{n}
      \edef\AMSa{\hexnumber\amsa@group}
      \makeatother
      \mathchardef\upi="0\UPM19
      \mathchardef\umu="0\UPM16
      \mathchardef\upartial="0\UPM40
      \mathchardef\leqslant="3\AMSa36
      \mathchardef\geqslant="3\AMSa3E

       \let\le=\leqslant

    \fi
  \fi
\fi % End of NFSS release 1

\ifnfsstwo
  \DeclareMathAlphabet{\mathbfit}{OT1}{cmr}{bx}{it}
  \SetMathAlphabet\mathbfit{bold}{OT1}{cmr}{bx}{it}
  \DeclareMathAlphabet{\mathbfss}{OT1}{cmss}{bx}{n}
  \SetMathAlphabet\mathbfss{bold}{OT1}{cmss}{bx}{n}
  \ifAMStwofonts
    \ifCUPmtlplainloaded \else
      \DeclareSymbolFont{UPM}{U}{eur}{m}{n}
      \SetSymbolFont{UPM}{bold}{U}{eur}{b}{n}
      \DeclareSymbolFont{AMSa}{U}{msa}{m}{n}
      \DeclareMathSymbol{\upi}{0}{UPM}{"19}
      \DeclareMathSymbol{\umu}{0}{UPM}{"16}
      \DeclareMathSymbol{\upartial}{0}{UPM}{"40}
      \DeclareMathSymbol{\leqslant}{3}{AMSa}{"36}
      \DeclareMathSymbol{\geqslant}{3}{AMSa}{"3E}

       \let\le=\leqslant

    \fi
  \fi
\fi % End of NFSS release 2

\ifCUPmtlplainloaded \else
  \ifAMStwofonts \else % If no AMS fonts
    \def\upi{\pi}
    \def\umu{\mu}
    \def\upartial{\partial}
  \fi
\fi

\title{Imaging Markarian 348 and its  water maser flare}
\title{MERLIN imaging of the maser flare in Markarian 348}
\author[E. Xanthopoulos \& A.M.S. Richards]
       {E. Xanthopoulos \& A.M.S. Richards \\ Jodrell Bank
        Observatory, University of Manchester, Macclesfield, Cheshire,
        SK11 9DL}
\date{ }

\pagerange{\pageref{firstpage}--\pageref{lastpage}}
\pubyear{}

\begin{document}
\bibliographystyle{mnras}

\maketitle

\label{firstpage}

\begin{abstract}
MERLIN images of Mrk 348 at 22~GHz show H$_{2}$O maser emission at
0.02 -- 0.11 Jy, within $\sim0.8$ pc of the nucleus. This is the first
direct confirmation that molecular material exists close to the
Seyfert~2 nucleus. Mrk 348 was observed in 2000 May one month after
\scite{Falcke00} first identified the maser in single-dish spectra.  The
peak maser flux density has increased about threefold.
The masing region is $\la0.6$ pc in radius. The flux
density of radio continuum emission from the core has been rising
for about 2 years.  The
maser-core separation is barely resolved but at the $3\sigma$
significance level they are not coincident along the line of sight.
The masers lie in the direction of the northern radio lobes and
probably emanate from material shocked by a jet with velocity close to
$c$. The correlation between the radio continuum increase and maser flare 
is explained as arising from high level nuclear activity through a 
common excitation mechanism  
although direct maser amplification of the core by masers tracing a 
Keplerian disc is not completely ruled out. 

\end{abstract}

\footnotetext{Contact e-mail: emily@jb.man.ac.uk, amsr@jb.man.ac.uk}
\section{Introduction}
\label{intro}

Water maser emission has been detected from galaxies up to $\sim100$
Mpc from Earth.  These supermasers are hundreds of times more luminous
than the brightest masers in star-forming regions in our own Galaxy
\cite{Cohen98}.  Such bright spectral lines are ideal targets for very
long baseline radio interferometry with micro-arcsec accuracy and with
 $\la1$~km~s$^{-1}$ velocity resolution.  This is the only way to image
distant galaxies directly on sub-pc scales.

H$_{2}$O supermasers are found exclusively in about 5\% of type 2
Seyfert and LINER galaxies \cite{Braatz97}.  The unified
scheme predicts that these active galactic nuclei (AGN) are obscured
by a molecular torus viewed edge on.  This provides high column
densities for maser amplification \cite{Kartje99}.  Such masers have
been used to measure the parameters of nuclear discs in Keplerian
rotation (e.g. \pcite{Miyoshi95}).
%(e.g. \pcite{Miyoshi95}, \pcite{Gallimore96}, A%\pcite{Greenhill97}).  
It is less obvious why the selection effect
applies to supermasers which appear to be associated with a jet
instead of  a circumnuclear disc (e.g. \pcite{Claussen98}).
%(e.g. \pcite{Claussen98}; \pcite{Trotter98}).  
These masers could
originate from interstellar material shocked by the jet
\cite{Elitzur92a}.

%Very recently 
\scite{Falcke00} reported the discovery of a very
luminous H$_{2}$O maser in Mrk 348 during a radio flare of the AGN.
Mrk 348 is a well studied Seyfert~2 at a redshift of 0.015
\cite{Huchra99}, with broad emission lines in polarized light
\cite{Miller90}.  However its host is an S0 galaxy at an angle of
inclination of only $\approx16^{\circ}$ \cite{Simkin87}.  Ground-based
\cite{Simpson96} and HST \cite{Falcke98} imaging have revealed a dust
lane crossing the nucleus. It has a high x-ray-absorbing column depth
of {\it N}$_{\rm H} = 10^{27.1}$ m$^{-2}$ towards the nucleus
\cite{Warwick89}.  These observations suggest the presence of an
obscuring torus in Mrk 348, but no molecular or H\,I absorption has been
detected so far (\pcite{Gallimore99}; \pcite{Falcke00}).

In this paper all velocities are given relative to the local standard of
rest ($V_{\rm LSR}$) in the radio convention.  The systemic $V_{\rm
LSR}$ of Mrk 348 is $4435\le V_{\rm sys}\le 4480$~km~s$^{-1}$ from
H{\sc i} emission line observations \cite{Bottinelli90} and imaging
\cite{Simkin87}.  We adopt $H_{0}$ = 75~km~s$^{-1}$ Mpc$^{-1}$, so
Mrk 348 is at a distance of $\sim60$ Mpc, where 1 mas = 0.29 pc.

What makes Mrk 348 stand out among Seyfert galaxies is its bright and
 variable inverted-spectrum radio nucleus \cite{Unger84}.  Jets at
 position angles (p.a.)  $~170^\circ$ and  $~30^\circ$ were observed with the EVN at 1.4
 GHz \cite{Neff83} and MERLIN at 5 GHz \cite{Unger84}.  \scite{Halkides97} first resolved the central
 part of Mrk 348, using the VLBA at 15 GHz, into two components
 separated by $\sim0.3$ pc at p.a. $\sim90^\circ$ to the larger 1.4
 GHz jet.  \scite{Ulvestad99} measured the sub-relativistic separation
 rate of these two components at the same frequency. They also noted a
 rise in continuum flux from 120 to 570~mJy between 1997.10 and
 1998.75.  Using MERLIN we can detect mJy radio continuum with a
 surface brightness $\ga5\times10^{4}$ K and locate bright masers with
 sub-pc relative positional accuracy. We use this to investigate the
 relationship between the maser and continuum flares and their origins
 in the core or in a nascent jet. 

\section{Observations and data reduction. }
\label{obs}
We observed Mrk~348 on 2000 May 2 using MERLIN, which
has a maximum baseline of 217~km, giving a beam size of 12~mas. 
In order to obtain images of Mrk 348 as rapidly as possible we
observed for a single 17 hr run in the maximum 16 MHz bandwidth, which
corresponds to a total velocity
width of $\sim200$~km~s$^{-1}$ and only covers  the red-shifted half of
the line seen by \scite{Falcke00} plus sufficient line-free channels
for continuum subtraction.  We observed Mrk~348 at a fixed frequency 
of
$\nu_{\rm o}$ = 21891.6~MHz alternately with phase reference source
J0057+3021 for 4 and 2 min respectively. 3C273, which
had a flux of 21.09 Jy at that time (Ter\"{a}sranta, private
communication), was observed for 40 min and used to set the flux
scale for all sources.  

Further data processing was performed using {\sc aips}
\cite{Greisen94}.  We applied the instrumental corrections from the
calibrators and the phase reference source solutions to Mrk~348.  We
then adjusted the data to a constant velocity, putting $V_{\rm LSR}$ =
4641.6~km~s$^{-1}$ in channel 30 of 60 usable channels, with a separation
of 3.37~km~s$^{-1}$. We averaged all data for Mrk~348 and made a {\sc
clean} map using natural weighting of the visibility data, which gave
a beam size of 31 mas $\times$ 21 mas. 

\begin{figure}
\vspace*{-4cm}
\hspace*{-0.8cm}
\rotatebox{-90}{
\resizebox{11cm}{!}{
\epsfbox{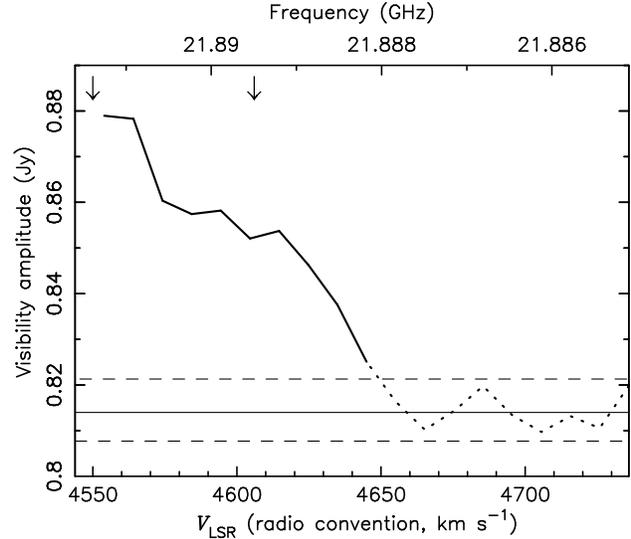}}}
\caption{The total flux density of Mrk~348 in each channel. 
The arrows mark the frequencies of peaks observed by Falcke et al. (2000).}
\label{xan-fig1.eps}
\end{figure}

The three {\sc clean} components ({\sc cc}) above $3\sigma_{\rm rms}$
from the initial map of Mrk~348 lay in a region 6 by 18 mas elongated
north-south. We used these as a model for phase self-calibration,
applied the solutions and examined the spectrum of Mrk~348.  The
complex visibilities were vector averaged in time channel by channel.
The flux density increased noticably with
increasing frequency if we used the more southerly {\sc cc} positions as the
origin of phase.  Fig.~\ref{xan-fig1.eps} shows the spectrum on the baselines to the Cambridge antenna
using the mean position of the maser emission (Table~\ref{position})
as the phase origin.  The low-frequency end of the band
appeared to be continuum only.  We used the first 4.25 MHz of data for phase
and  amplitude self-calibration and applied the same corrections to the 
line-only data as we did for the continuum data. 

The multi-channel Mrk~348 data were averaged over every 0.75~MHz and
Fourier transformed to make a 20-channel total emission dirty data
cube.  The average of the first 4.50~MHz was subtracted pixel by pixel
from the whole cube to leave a line-only dirty data cube which was
{\sc clean}ed to produce a {\sc clean} line-only cube. The line-only
dirty data cube was subtracted from the total emission dirty data cube
and the result was {\sc clean}ed to give a continuum-only cube. We
also made a {\sc clean} total emission cube.

We fitted 2D Gaussian components to emission above $3\sigma_{\rm rms}$
in each channel of every {\sc clean} cube to determine the position
 and peak flux $S_{\rm p}$.  We also measured the position
and flux density of components fitted to the total flux from Mrk~348
mapped using 15-MHz band-width prior to self-calibration and the
4.25-MHz bandwidth self-calibrated continuum map.  The brightest
continuum components was resolved after deconvolving the beam, so we
could measure its FWHM $s$ and the integrated flux $S_{\rm i}$. The
position and component size
uncertainties ($\sigma_{\rm pos}$, $\sigma_{\rm s}$)
 are proportional to the beam-width divided by the
signal-to-noise ratio allowing for the sparse baseline coverage
(\pcite{Condon98}; \pcite{Richards99}).

 All maps are presented in
total intensity.  The flux scale should have
$<10\%$ error but MERLIN only has 5 antennas operating at 22 GHz, and
the sparse $uv$ coverage means it is possible for extended continuum
emission to appear brighter and more compact than it actually is,
although the peak positions should be accurate. Moreover as only one
side of the bandpass is line-free, errors in baseline subtraction or
bandpass calibration
 may affect the maser flux measurements. The absolute position
accuracy is $\sim50$ mas, mostly due to uncertainty in the position of 
J0057+3021 \cite{Wilkinson98}.

%We have found that for MERLIN observations using $N$ antennas
%including Cambridge this must be multiplied by $\sqrt{N/2}$
%\cite{Richards99}.  This is adequate for sources with $10
%\sigma_{\rm rms}\la S_{\rm p} \la 150 \sigma_{\rm rms}$, observed
%with good hour angle coverage, which can be well approximated by an
%elliptical Gaussian component (e.g. the Mrk 348 continuum
%emission). For faint ($S_{\rm p}\la10\sigma_{\rm rms}$) unresolved
%sources such as the maser emission the fitting uncertainty must also
%be taken into account, satisfactorily achieved by setting $\sigma_{\rm
%x}$ and $\sigma_{\rm y}$ to $3\times$ the {\sc jmfit} errors.

\section{Results}
\label{results}

\begin{figure}
\resizebox{8.6cm}{!}{\epsfbox{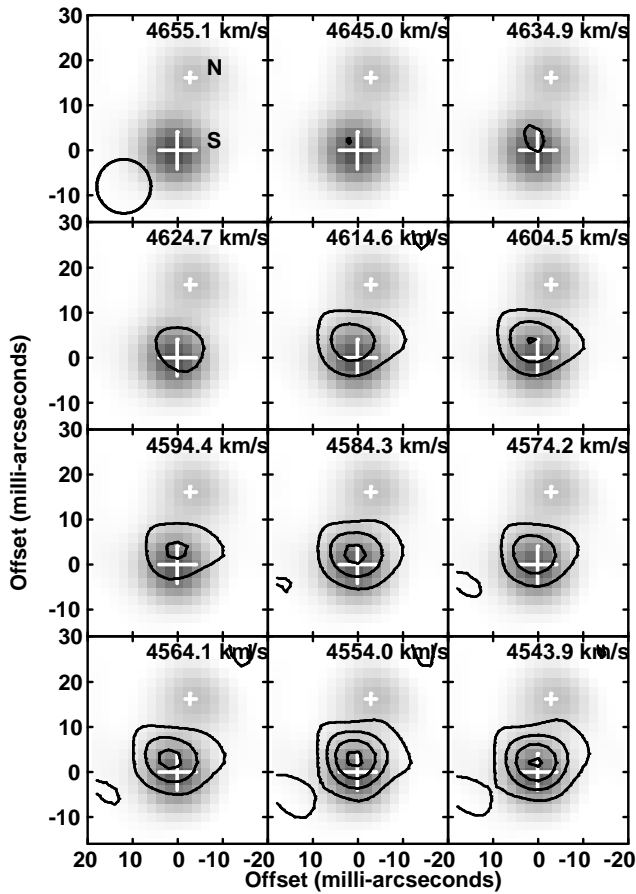}}
\caption{H$_{2}$O maser emission from Mrk~348 is shown by the
contours, at (--1, 1, 2, 3, 4...)$\times20$~mJy~beam$^{-1}$.  The grey
scale shows continuum emission from 0 to 1 Jy~beam$^{-1}$.  The
circular restoring beam of 12 mas FWHM is shown in the top left
panel. 
 The white crosses show the continuum peaks {\bf \sf N} and {\bf \sf
S}, size proportional to the peak flux.   The axes are labelled in mas offset from {\bf {\sf S}}.
Each panel is labelled with the $V_{\rm LSR}$
(radio convention).}
\label{xan-fig2.eps}
\end{figure}

A single patch of maser emission was detected in 11 channel maps of
the line-only data cube from 4543.8 to 4655.0~km~s$^{-1}$, shown by
the contours in Fig.~\ref{xan-fig2.eps}.  These are in linear
multiples of $3\sigma_{\rm rms}$.  The grey-scale shows the
continuum-only emission above $3\sigma_{\rm rms}$. The continuum peaks
are marked {\bf \sf N} and {\bf \sf S} and the white crosses show
their positions, which agree in each channel to within 0.1 mas. 

The positions and flux densities of the peaks are given in
Table~\ref{position}.  The peak of the maser emission is consistently
offset from the continuum peak; its mean position {\bf \sf \={M}} is
$2.7\pm0.7$ mas north of {\bf \sf S}.  The total angular size of the
maser region is $\le3$ mas and the individual components are
unresolved.  There is no significant systematic angular
separation-velocity gradient greater than 2~mas in
111~km~s$^{-1}$. 
{\bf \sf S} appears to be resolved, and was fitted
with a component of $2.0\pm0.2$ mas FWHM and an integrated flux of 814
mJy giving a brightness temperature of 551 -- 824$\times10^6$ K. {\bf
\sf N} is at a position angle of $-11^{\circ}\pm2^{\circ}$ with
respect to {\bf \sf S}.

\begin{table*}
\caption{The positions 
and peak intensities  $S_{\rm p}$ of
the 22 GHz continuum and maser peaks in Mrk 348 {\bf \sf N}, {\bf \sf S} and {\bf \sf \a={M}}.} 
\label{position}
\begin{tabular}{lccccc}
Peak  & R.A. & Dec.  & $\sigma_{\rm pos}$  (mas) &$S_{\rm p}$ (mJy  beam$^{-1}$)      	&$\sigma_{\rm rms}$	(mJy  beam$^{-1}$)\\
\hline	  	 		  	 		        		  		 		    			 
{\bf \sf S} & ~00\h~48\m~47\fs14575 & +31\degr~57\arcmin~25\farcs1128 &   0.2        &  744            	&	    6		\\
{\bf \sf N} & ~00\h~48\m~47\fs14552 & +31\degr~57\arcmin~25\farcs1470 &   0.4         &   239          	&     6\\
{\bf \sf \a={M}}  & ~00\h~48\m~47\fs14580 & +31\degr~57\arcmin~25\farcs1155  &  0.7       &    24 -- 107     	&    8	\\
\hline
\end{tabular}
\end{table*}

\begin{figure}
\vspace*{-4cm}
\hspace*{-0.5cm}
\rotatebox{-90}{
\resizebox{10cm}{!}{
\epsfbox{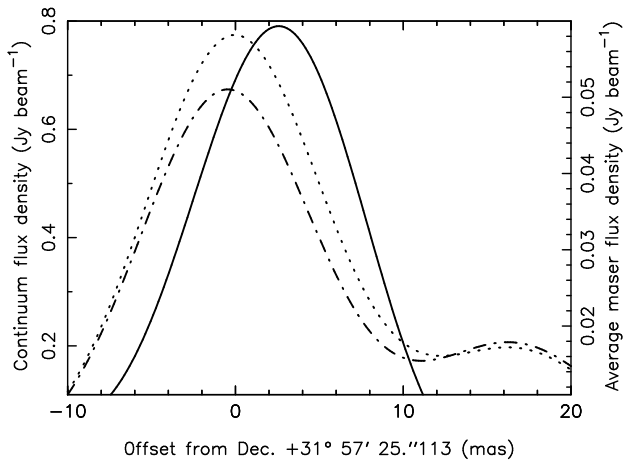}}}
\caption{Flux density as a function of position along slices through
the maser and continuum peaks.  The dot-dashed line shows the slice
taken from the average of continuum-only channel maps.  The dotted
line shows the average of channel maps containing maser and continuum
emisson. The solid line shows the averaged maser emission in these
channels after continuum subtraction, note the different flux scale.}
\label{xan-fig3.eps}
\end{figure}

  The maps of continuum-only channels from 21884.35 to 21888.60 MHz,
the continuum+line channels from 21890.85 to 21899.10 MHz and
line-only channels from 21890.85 to 21899.10 MHz were averaged from
their respective data cubes.  Flux measurements along slices through
these averaged maps, at constant R.A., 2~mas wide, intersecting the
position of {\bf \sf S}, are plotted in Fig.~\ref{xan-fig3.eps}.  This
shows the offset of the maser peak from the brighter continuum peak.

%\subsection{Absolute position accuracy}

%Table~\ref{refpos} lists the position of the brightest continuum
%component of Mrk~348 observed using MERLIN, measured after applying
%phase-reference source solutions only, compared with published
%positions.  

%The positions of the brightest component 
%measured by MERLIN at different frequencies differ by up to 11 mas,
%and are $\sim40$ mas offset from the position given by
%\scite{Wilkinson98}.  
%This discrepancy is within the known position
%errors, and it is also likely that the source does not have a uniform
%spectral index within the region covered by the beam, further
%contributing to the shift of the peak position at difference
%frequencies.  The position given by \scite{Ulvestad99} is $\sim90$ mas
%away, but they do not specify the phase reference source used. 

\section{Discussion}

%We have observed the newly discovered megamaser Mrk~348 with MERLIN at
%22 GHz in order to study the  masers at higher resolution, test
%whether there is a correlation between the continuum and maser flux
%density and investigate their origin in the core or jet, which could
%give us some more insight to the maser excitation mechanism.  Our
%results are based on observations that have a total velocity width of
%$\sim200$~km~s$^{-1}$ which covers half, the redshifted part, of the
%asymmetric line discovered by \scite{Falcke00}.

These observations are the first direct detection of molecular
material in the nuclear region of Mrk 348.  The existence of H$_{2}$O
implies shielding by a dusty medium with a column density $\ga10^{26}
- 10^{27}$ m$^{-2}$ and the conditions required for population
inversion of the maser include a gas number densitity $\sim10^{14} -
10^{16}$ m$^{-3}$, a fractional abundance of H$_{2}$O $\sim10^{-5} -
10^{-4}$ and a temperature $>250$ K \cite{Kartje99}.  Conditions can
be more tightly constrained depending on the association of the masers
with a disc or a jet.

%We assume that the brightest unresolved
%22-GHz continuum component {\bf \sf N} contains the core and 2-mas
%northern jet at position angle $15^{\circ}$ found by
%\scite{Ulvestad99} at 15~GHz. {\bf \sf S} is at a similar direction to
%the jet detected at 1.4 and 5~GHz at position angle $163^{\circ}$ --
%$170^{\circ}$ (\pcite{Neff83}; \pcite{Unger84}). 

We assume that the brightest 
22-GHz continuum component {\bf \sf S} contains the core and 2-mas
northern jet at position angle $15^{\circ}$ found by
\scite{Ulvestad99} at 15~GHz. {\bf \sf N} is in a similar direction to
the more extended jet detected at 1.4 and 5~GHz 
(\pcite{Neff83}; \pcite{Unger84}).

We consider three possible models for the relationship between the
maser flare and the continuum flare:
\begin{enumerate}
\item{The masers trace a small warped disc and the
maser emission follows a Keplerian rotation law, and the masers directly
amplify the continuum emission.}
\item{The masers are unsaturated and  lie in a
symmetric Keplerian nuclear disc. Perturbation of the disc
creates spiral shocks \cite{Maoz98} that are also responsible for the continuum flare.}
\item{The radio continuum flare is associated with the ejection of
material in the direction of the northern jet. The masers arise from the
ISM where it is shocked by the jet.}
\end{enumerate}
 Below we expand on the models and explain why
we prefer or reject each.

The ratios of the peak fluxes April:May are 27:77 at 4604~km~s$^{-1}$
and 26:107 at 4544~km~s$^{-1}$, a 3 -- 4-fold increase.  Although
further observations are essential in order to pin down the source
timescale, such rapid variability may suggest that the individual
masing clouds are less than 1 light-month (0.025 pc) in diameter.
This would then be about half the size of the 15-GHz core
\cite{Ulvestad99} and so the maser emission that is beamed in our
direction is compact enough to originate from maser clouds that
overlap our line of sight to the core.  However the two arrows in
Fig.~\ref{xan-fig1.eps} indicate the velocities of the peaks that
\scite{Falcke00} observed and which are close to the peaks in our
data, but $\sim100$~km~s$^{-1}$ more redshifted than $V_{\rm sys}$.
\scite{Falcke00} did not detect any maser emission at  $V_{\rm LSR}<V_{\rm sys}$.
So explanation (1) is inconsistent with a Keplerian disc unless the
masing material is infalling.  Moreover the maser peak is consistently
misaligned with {\bf \sf S} in every channel in
Fig.~\ref{xan-fig2.eps}.  The $3\sigma$ position error boxes for {\bf
\sf \={M}} and {\bf \sf S} (Table~\ref{position}) are too close to
rule out direct maser amplification of the continuum peak, but we
consider other geometries are more probable.

%They detected maser emission between velocities
%close to $V_{\rm sys}$ and 220~km~s$^{-1}$ more red-shifted.
\scite{Maoz98} found a tendency for the red-shifted emission from disc
supermasers around AGN to be brighter and explained this using a model
of spiral shocks within a Keplerian disc.  This mechanism requires
perturbation of the disc, which could be connected to the event which
caused the continuum flare.  Such a disc should be symmetric about the
nucleus with a well-defined position-velocity gradient.  If it is
nearly edge-on the rotation velocity is $V_{\rm max}-V_{\rm sys}
\approx 200$~km~s$^{-1}$.  The MERLIN results show the velocity
gradient is $\le2$ mas in 111~km~s<$^{-1}$ corresponding to a disc
radius $\la 1.1$ pc.  
{\bf \sf \={M}} is $0.8\pm0.2$ pc north of {\bf \sf S}.  This is
consistent with the $\la2$ yr time-lag between detection of the maser
and continuum flares which implies a separation of $\la0.6$ pc
\cite{Falcke00}.  This implies that if the masers  lie in a symmetric
disc it is  elongated north-south, parallel to the radio jets.  This
is unlikely and so model (2) is also improbable.

%The mass $M$ enclosed by such a disc can be crudely estimated using Kepler's
%second law, in convenient units: 
%\begin{equation} 
%\frac{M}{\rm M}_{\odot} = 231 \frac{r_{\rm d}}{\rm pc}
%\left(\frac{V_{\rm d}}{\rm km\/ s^{-1}}\right)^2 
%\end{equation}.

Model (3), in which the masers originate in a shock produced by the
northern jet, is possibly due to the misalignment of the nuclear disc
with respect to the host galaxy.  If this material was ejected from
the core when the continuum flare commenced $\la$ 2 years prior to the
maser flare this implies speeds near $c$ .  The only previous
detection of a relativistc Seyfert jet is in III Zw 2
\cite{Brunthaler00}. However \scite{Ulvestad99} measured a jet speed
of $\sim0.07c$ for Mrk 348 on similar scales suggesting the jet power
is rapidly dissipated once it reaches the shocked region.

If the maser flare is not simply direclty
amplifying the  radio continuum flare this 
indicates that there is some common excitation effect, 
possibly some sort of high level nuclear activity.
If the masers the masers are found along the radio jets, as is the case 
for NGC 1052 (\pcite{Claussen98}), this mechanism for the 
 correlation between the evolution of the maser flare and 
the radio flare is an important tool to study the 
jet-ISM interactions. 

We have observed the newly discovered megamaser Mrk~348 with MERLIN at
22 GHz in order to study the  masers at higher resolution, test
whether there is a correlation between the continuum and maser flux
density and investigate their origin in the core or jet, which could
give us some more insight to the maser excitation mechanism. 
Further imaging observations of the whole line will distinguish
between these possibilities and elucidate the origins of the flare.

\section{Acknowledgements}

We would like to thank our referee Dr. Huib Jan Van Langevelde for 
many constructive comments.
MERLIN is the Multi Element Radio Linked Interfermometer Network, a
national facility operated by the University of Manchester at Jodrell
Bank Observatory on behalf of PPARC.  We thank the MERLIN
staff for performing the observations, and Peter Thomasson, Simon
Garrington and Tom Muxlow for useful discussions.  We are 
grateful to Phil Diamond,  Alan Pedlar and Alison Peck for  very helpful
contributions to this paper.
This research was supported by European Commission, TMR Programme,
Research Network Contract ERBFMRXCT96-0034 ``CERES". 

%\bibliography{/scratch/aife_1/amsr/scathach_1/amsr/SAVE/H2Opaper/cse}
\bibliography{mrk348_june2001.bbl}

\bsp

\label{lastpage}

\end{document}